\documentclass[sigconf]{acmart}

\usepackage[utf8]{inputenc}

\AtBeginDocument{%
  \providecommand\BibTeX{{%
    \normalfont B\kern-0.5em{\scshape i\kern-0.25em b}\kern-0.8em\TeX}}}

\setcopyright{rightsretained}

\usepackage{csquotes}
\usepackage{amsmath}
\usepackage{booktabs}
\usepackage{graphicx}

\begin{document}

\title[Machine Learning Uncertainty as a Design Material]{Machine Learning Uncertainty as a Design Material: \\ A Post-Phenomenological Inquiry}

\author{Jesse Josua Benjamin}
\email{j.j.benjamin@utwente.nl}
\affiliation{%
  \institution{Department of Philosophy \\ University of Twente}
  \city{Enschede}
  \country{Netherlands}
}
\author{Arne Berger}
\email{arne.berger@hs-anhalt.de}
\affiliation{%
  \institution{ Computer Science and Languages \\ Anhalt University of Applied Sciences}
  \city{Koethen}
  \country{Germany}
}
\author{Nick Merrill}
\email{ffff@berkeley.edu}
\affiliation{%
  \institution{Center for Long-Term Cybersecurity \\ University of California, Berkeley}
  \city{Berkeley}
  \country{United States}
}
\author{James Pierce}
\email{jjpierce@uw.edu}
\affiliation{%
  \institution{School of Art + Art History + Design \\ University of Washington}
  \city{Seattle}
  \country{United States}
}
\renewcommand{\shortauthors}{Benjamin, et al.}

\begin{abstract}
Design research is important for understanding and interrogating how emerging technologies shape human experience. However, design research with Machine Learning (ML) is relatively underdeveloped. Crucially, designers have not found a grasp on ML uncertainty as a design opportunity rather than an obstacle. The technical literature points to data and model uncertainties as two main properties of ML. Through post-phenomenology, we position uncertainty as one defining material attribute of ML processes which mediate human experience. To understand ML uncertainty as a design material, we investigate four design research case studies involving ML. We derive three provocative concepts: \textit{thingly uncertainty}: ML-driven artefacts have uncertain, variable relations to their environments; \textit{pattern leakage}: ML uncertainty can lead to patterns shaping the world they are meant to represent; and \textit{futures creep}: ML technologies texture human relations to time with uncertainty. Finally, we outline design research trajectories and sketch a post-phenomenological approach to human-ML relations.
\end{abstract}

\ccsdesc[500]{Human-centered computing~HCI theory, concepts and models}

\copyrightyear{2021} 
\acmYear{2021} 
\acmConference[CHI '21]{CHI Conference on Human Factors in Computing Systems}{May 8--13, 2021}{Yokohama, Japan}
\acmBooktitle{CHI Conference on Human Factors in Computing Systems (CHI '21), May 8--13, 2021, Yokohama, Japan}\acmDOI{10.1145/3411764.3445481}
\acmISBN{978-1-4503-8096-6/21/05}

\keywords{post-phenomenology, machine learning, design research, thingly uncertainty, horizonal relations}


\maketitle

\section{Introduction}
Just as prior advances in computing technology have led to design researchers approaching algorithms, software interfaces, sensors, and actuators as \textit{design materials} \cite{redstrom_technology_2005, lim_interactivity_2011}, similar developments are happening with Machine Learning (ML) technologies (cf. \cite{hebron_machine_2016,leahu_ontological_2016,dove_ux_2017,yang_mapping_2018}). Recent advances in this regard have resulted in guidelines from computing corporations for curating ML results, improving user experience design and Explainable Artificial Intelligence (XAI) (e.g., \cite{amershi_guidelines_2019,google_pair_people_2019,yang_investigating_2018,liao_questioning_2020}). In light of the increasing ubiquity and technical opacity of ML, design research methodologies such as Research-through-Design, Speculative Design or Design Fiction are urgently needed in this space to develop better tools grounded in a rich and socially-situated understanding of how ML shapes everyday life. 

We argue that design research is well-posed to describe, explore and reflect on how the ``statistical intelligence'' \cite{dove_ux_2017} of ML decision-making bleeds into the intimate human experience of lifeworlds; and to productively engage with emerging personal, societal and ethical issues. At the same time, we observe that design research struggles to engage ML as a material for design, as the probabilistic inference of models from data patterns withdraws from being present-at-hand. A major focus of prior work is the technical opacity of ML. However, just as ML-driven systems are difficult to interpret and exhibit emergent and oftentimes unpredictable behavior \cite{burrell_how_2016}, the outputs they generate are inherently characterized by uncertainty from data noise and model variance \cite{kendall_what_2017}. 

ML uncertainty is a problem for HCI design research, because the field has not yet framed it as a design material to assess, design, and reflect novel applications, objects and services that build on ML. On the contrary, prior work, for example in the area of XAI, is often concerned with using design methods to explain, rather than utilize, ML uncertainty. We propose that approaching ML uncertainty as a specific material property of ML can help design research consider novel spaces for design intervention and interaction.

In this paper, we investigate ML uncertainty from a post-pheno-menological perspective to develop a conceptual vocabulary for engaging ML uncertainty in designerly ways. The principal argument of post-phenomenology is that technologies actively \textit{mediate} human relations to the world \cite{ihde_technology_1990}: technologies are neither completely deterministic (technological determinism) but neither are they neutral tools (technological instrumentalism). Instead, human-technology relations are co-constitutive, with artefacts shaping how humans as specific subjectivities relate to the world \cite{verbeek_what_2006}. In this view, design is \textit{doing} philosophy ``by other means'' \cite{verbeek_beyond_2015}, and design materials are those involved in shaping technological mediation. We follow Hauser et al.’s framing of Research-through-Design (RtD) as post-phenomenological practice \cite{hauser_annotated_2018}, and expand this view to further design research methodologies and the specific topic of ML. We thereby also continue prior work from the HCI community on using and extending the post-phenomenological framework (cf.~\cite{wakkary_morse_2017,hauser_deployments_2018,pierce_smart_2019}). Specifically, we investigate four design research case studies using a post-phenomenological lens, and develop provocative concepts for ML uncertainty similar to ``strong concepts''~\cite{hook_strong_2012} or robust ``annotations''~\cite{gaver_annotated_2012,hauser_annotated_2018} in HCI. In this, we take a modest step forward in making ML uncertainty tangible as a design material for HCI design researchers. We argue that uncertainty is the material expression (cf. \cite{redstrom_technology_2005}) of ML \textit{decision-making}. While there are reasonable engineering incentives to minimize uncertainty in many use cases, uncertainty constitutes a fundamental attribute of any ML-driven system. Uncertainty offers a representation of the `fault lines' of ML decision-making, not only as a negative attribute of solutions but as simply part-and-parcel of ML technologies. As such, designerly making with uncertainty offers an opportunity to design artefacts and scenarios that attribute or more fully exploit the characteristics of ML decision-making. Our contributions consist of provocative, conceptual shorthands for ML uncertainty. We encourage designers to not see ML uncertainty as “to be explained away,” but rather as generative of particular relations that can be designed for. At the same time, we also show how emerging ML applications are not readily accounted for with prior human-technology relation concepts.

We provide three provocative concepts that build on ML uncertainty as a source for future research in HCI, design research and philosophy of technology communities. 
As a general concept, we propose \textit{thingly uncertainty} to capture the capacity of ML-driven artefacts to be uncertain about the world, and thereby generating and adapting to a wide continuum of relations to other things, their datafied environment and people. 
We furthermore distinguish two specific concepts. First, we propose \textit{pattern leakage} for describing how ML models become generative of patterns which enter and alter the everyday. Pattern leakage describes how ML models alter the world they seek to represent. 
Second, we propose \textit{futures creep} to denote the often subtle transformations of the present by ML predictions; changing human relations to time by injecting probabilistic events such as climate change predictions into the direct perception of the present. 
Based on our derived concepts, we propose how design research can engage with ML uncertainty, and furthermore suggest \textit{horizonal relations} as a post-phenomenological research trajectory into the specifics of how ML capacities and human experience co-extend and overlap. Among the contributions of this paper, we (1) construct a post-phenomenological lens on ML based on related work; (2) analyze four case studies from different design research methodologies through this lens to discern ML uncertainty as a design material; (3) present \textit{thingly uncertainty}, \textit{pattern leakage} and \textit{futures creep} as provocative concepts for future work; and (4) propose \textit{horizonal relations} as a distinct human-technology relation, and lay out research trajectories for investigation.

\section{Background}
In the following section we (1) provide a brief introduction to ML uncertainty; (2) frame how design research can pragmatically and critically engage with emerging technologies; (3) argue that post-phenomenology is a promising lens to make ML uncertainty graspable in design research.

\subsection{ML Technologies between Opacity, Interpretability and Uncertainty}
Many of the diverse algorithmic techniques common in ML today stem from cybernetics. McCulloch and Pitt's notion of neuronal activity \cite{mcculloch_logical_1943} and Rosenblatt's \textit{Perceptron} as a probabilistic model for learning and remembering information about the environment \cite{rosenblatt_perceptron_1958} are in direct connection to today's advances in deep learning, while Wiener's concept of negative feedback \cite{wiener_cybernetics_1961} can still be seen as the general principle which make ostensibly novel technologies tick. And like cybernetics, today's ML technologies are reliant on probabilistic techniques as a computational means for the ``taming of chance'' \cite{hacking_taming_1990}. The deployment of increasingly powerful probabilistic techniques (e.g., expectation-maximization, gradient descent, backpropagation) capable of adapting models to large datasets in real-world settings has also come at the cost of opacity, and become generative of types of uncertainty originating from \textit{within} technological deployments themselves.

In general, ML algorithms operate according to the principle of ``insight through opacity'' \cite{mcquillan_data_2018}, using specific probabilistic techniques to infer a model that describes statistical patterns in datasets \cite{bishop_pattern_2006}, which approximates some assumed real-world functional relationship \cite{goodfellow_deep_2016}. The insight, i.e. a model that can detect significant patterns in relevant data (e.g., does this image show tumors?), depends on opacity: as each variable such as a pixel is computed as a vector in relation to all other variables, the combined dimensionality of vectors exceeds those that humans can intuit. In short, the often ``unreasonable effectiveness'' \cite{halevy_unreasonable_2009} of current ML implementations comes at the cost of understanding how exactly insights were arrived at. From within ML research, the challenge of opacity is one of the most urgent topics of research, showing in fields such as XAI or ML interpretability research. A common approach for ML research is to deploy algorithmic methods, so-called interpretability techniques \cite{lipton_mythos_2016,miller_explanation_2017}, to extract information from ML pipelines in order to explain outputs via textual or visual explanations. While this research area is fundamentally oriented at experts, its focus on actual ML technologies offers up a catalogue of properties potentially of interest to design research. Given ML's epistemological as well as technical origins in cybernetics, we are specifically interested in how uncertainty is dealt with in light of the contemporary insight-through-opacity approach.

In ML research, the aim of dealing with uncertainty takes on a distinctly \textit{material} grounding: given the reliance on probabilistic techniques, uncertainty is not only a human disposition but part and parcel of ML technologies. ML researchers frequently distinguish between two types of uncertainty which are related to data (both training and input) and models respectively \cite{kiureghian_aleatory_2009, fox_distinguishing_2011, kendall_what_2017}. The former, ``aleatoric'' uncertainty, can be framed as `noise': incoming signals in training or real-world deployments are inevitably impure and may affect the performance of ML algorithms. The latter, ``epistemic'' uncertainty, refers to the complex questions surrounding the `fit' of a generated ML model. Inferred models embody, firstly, only one way of describing patterns from a given dataset, and given the insight-through-opacity approach the relationship to unknown models is uncertain. Secondly, an inferred model may also \textit{generate} additional uncertainty when deployed ``out-of-data'', i.e. in settings different to training environments. Data and model uncertainty frequently feature in ML interpretability research. It is important to note that both are \textit{computable} depending on the given ML deployment. Hohman et al., for example, deploy a technology probe for ML experts featuring various data visualizations in one interface, which includes a ``regions-of-error'' technique showing the model uncertainty of predictions \cite{hohman_gamut:_2019}. Similarly, Kinkeldey et al. use a landscape metaphor in a cluster visualization, indicating through a grey-scale topography how certain the clustering model is about the membership of each individual point by their location in ``peaks or slopes'' \cite{kinkeldey_towards_2019}. Concerning data uncertainty, Kendall and Gal note that in image segmentation using deep learning, noise affects the boundaries surrounding objects \cite{kendall_what_2017}. Similarly, Kwon et al. note that data uncertainty in brain lesion detection with neural networks manifests around affected brain regions \cite{kwon_uncertainty_2018}. 

On a general level, we therefore consider data uncertainty to manifest with the \textit{objects} of ML decision-making (e.g., data as images, strings, vectors), whereas model uncertainty concerns the \textit{mode} of ML decision-making (e.g., clustering, classifying, predicting). While the technical fields take an understandably solutionist stance on ML uncertainty by attempting to either minimize or explain it, we argue that the notion of computational uncertainty as a part-and-parcel property of ML promises opening actual ML technologies to designerly research.

\subsection{Design Research in ML and Emerging Technologies}
Uncertainty, understood in an everyday sense as ambiguity or chance, is a well-known resource in design research (cf.~\cite{gaver_cultural_2004}). As an alternative to quantitative studies, design research focuses on unearthing, exploring and understanding individual and situated encounters of people with technology that are often based on uncertainty and indeterminacy. Particularly ‘third-wave HCI’ \cite{harrison_making_2011} methodologies such as Research-through-Design, Design Fiction or Speculative Design have engaged this trajectory, foregrounding how technological artefacts are not merely solutions to discrete problems but rather embody open-ended, contextually dependent questions. Yet, design research has predominantly focused on either using ML \textit{for} design, or using design \textit{for} ML. With regards to the former, for example, Yang et al. have discussed how ML is engaged by design practitioners and researchers \cite{yang_mapping_2018,yang_grounding_2018,yang_investigating_2018} to improve user experience through adaptation or personalization. They found that while practitioners and researchers are enthusiastic about using ML, design research so far lacks distinct methods that engage ML as a design material. Similarly, Dove et al., conclude that integrative prototyping methods reflecting both ``ML statistical intelligence and human common sense intelligence’’ \cite{dove_ux_2017} are missing in the field. However, in contrast to this research focus on ML \textit{for} design, we argue that more fundamental research is needed into how, and to what end, ML technologies can be a design material in their own right. Similarly, in HCI approaches to XAI, researchers envision design research methods for improving a given system's explainability; by e.g. developing explainability scenarios \cite{wolf_explainability_2019,andres_scenario-based_2020} or conceptualizing contextually sensitive questions \cite{liao_questioning_2020,wang_designing_2019} for various stakeholders. We see a similar, if inverted, limitation in this use of design \textit{for} ML: in XAI, ML technologies tend to become the target (for e.g. designing explanations), not the material, of design research methods. We therefore next consider critical design research projects that, while not always directly addressing ML technologies, deal with uncertainties of complex design materials.

For example, Merrill et al. found that lay people believe different types of biosensors can reveal much more, or much less, than they actually can \cite{merrill_sensing_2019}. Others have explored how people speculate on smart things for the home, based on the fluidity and diversity of the values they attribute to their individual homes. This unraveled the idiosyncrasies and situatedness of potential future smart things, and highlights that these are bound to individual experiences with uncertainty \cite{berger_inflatable_2019,oogjes_designing_2018}. Pierce’s design-led inquiry into smart camera systems \cite{pierce_smart_2019} investigates the ways in which smart camera systems opaquely embody specific relations to the world--that is, functions which may be concealed due to undesired effects (e.g., distrust, failure). Redstr{\"o}m and Wiltse further interrogate how user interactions are tied to infrastructural functions, and outline how ``surface-level simplicity'' of interactions such as pressing play in Spotify belie ``dynamic, sophisticated, and hidden backend complexity''~\cite{redstrom_press_2015}.

This brief overview of critical design research projects shows that in general, situated yet concealed design materials are nothing new, and opacity and its often uncertain effects are a recurring theme across application areas. However, we observe that few critical design research projects deal with specific rather than generic ML technologies. We hypothesize that design research currently lacks a conceptual grasp on material properties that characterize ML technologies' inference of models from data and decision-making (e.g., prediction, classification). Due to their relationship with input, inference and output of ML technologies, we propose that computational model and data uncertainty from the technical fields discussed above are promising candidates. That is, rather than a human-centered notion of uncertainty, we propose that the technical framing of uncertainty promises a stronger foothold on the ML design space. However, it is unclear how ML uncertainty can be framed specifically for design research. For example, Hemmentt et al. call for artistic, designerly practices of revealing the ``distortions in the ways in which algorithms make sense of the world'' \cite{hemment_experiential_2019}; yet do not outline how this may relate to existing design research methodologies as well as actual ML technologies. In the following, we discuss how the philosophical framework of post-phenomenology offers a starting point to expand on this gap.

\subsection{Post-Phenomenology and Sketching Human-ML Relations}\label{ssc:hmlr}
Post-phenomenology is an empirical-analytical framework in philosophy of technology. In this view, technological artefacts shape human perception and action by mediating the world in specific ways \cite{ihde_technology_1990}. The framework has become particularly influential in HCI design research, as it foregrounds the responsibilities and ethico-political stakes of designing technological artefacts—-surfacing how technologies co-constitute our relations to specific ‘slices’ of the world, and how this relation makes us who we are \cite{frauenberger_entanglement_2019,hauser_annotated_2018,verbeek_beyond_2015}. 

Below, we outline established concepts of post-phenomenology in the form of their basic relational schemata, which allow empirical-analytical research to probe their objects of study for the roles that \texttt{human}, \texttt{technology} and \texttt{world} play. 

The schemata use established notations for simple \textit{connections} between entities (\texttt{-}); \textit{interpretation} of one by the other (\texttt{→}); being experienced together (\texttt{()}); being in the \textit{background} (\texttt{/}) of another entity; or being already \textit{thematized} (i.e., meaningfully focused  \cite{heidegger_being_2010}) in some way \texttt{([])} before being experienced. We then argue that post-phenomenology is a well-suited framework to describe, analyze, and interpret ML as a technology with active yet concealed relations to the world.

\subsubsection{Technological Mediation:} 
Human ways of perceiving and acting in the world are shaped by technological artefacts, co-constitu-ting who relates to what/whom in which way. Depending on the \texttt{technology} (e.g., an ultrasound scanner), ways of perceiving phenomena in the \texttt{world} are shaped (e.g., the fetus-as-patient, the womb-as-potentially-dangerous-enclosure), and actions are invited or inhibited (e.g., decisions on fetal care), for a specific \texttt{human} subjectivity (e.g., the non-pregnant parent-as-caretaker) \cite{verbeek_what_2006}. The general schema for technological mediation is:

\begin{center}
\small
\texttt{Human\,--\,Technology\,--\,World}
\end{center}

\subsubsection{Human-Technology Relations:}\label{htr}
Post-phenomenology studies technological mediation through human-technology relations (\autoref{tableHTR}), the structures of the empirical settings in which humans and technologies encounter each other, and how such encounters shape how humans relate to the world \cite{rosenberger_postphenomenological_2015}. 

Ihde initially proposed four types of human-technology relations \cite{ihde_technology_1990}. In \textit{embodiment relations}, technologies become an inseparable part of human bodily-perceptual experience. Wearing glasses shapes our experience of the world, not of the glasses themselves. In \textit{hermeneutic relations}, technologies make the world legible in a specific way. Reading a map mediates the world as a grid; a thermometer translates heat phenomena onto a legible scale. More overt, in \textit{alterity relations} technologies are seen as a quasi-Other. For example, an ATM, toy or robot are interacted with as if they have human-like intentions. In \textit{background relations}, technologies merge with the background of experience, yet ``texture'' (i.e., generate an atmosphere for) that experience. Central heating or thermostats are absent presences in the experience of our home, but are rarely interacted with directly. Verbeek further expanded these relations to reflect the more subtle and also radical mediation by emerging technologies. In \textit{immersion relations}, for instance, technologies such as augmented reality glasses or ambient intelligences merge with the environment, shaping how social relations can be enacted within them \cite{verbeek_designing_2015}. More radically, in \textit{cyborg relations}, technologies such as microchips or pacemakers merge with the body, becoming indistinguishable in direct experience \cite{verbeek_cyborg_2008}. And lastly, in \textit{composite relations}, technologies such as extremely long-exposure photography or computational imaging make things experienceable that have no direct correlation to ordinary human modes of perceiving space and time. Instead, composite relations mediate a ``reality that can only be experienced by technologies'' \cite{verbeek_cyborg_2008}. 

\begin{table}[h!]
\resizebox{1\columnwidth}{!}{%
\begin{tabular}{@{}lll@{}}
\toprule
Relation & Schema & Examples \\ \midrule
Embodiment & \texttt{(I-Technology)World} & Glasses, Cane \\
Hermeneutic & \texttt{I} → \texttt{(Technology - World)} & Thermometers, Maps \\
Alterity & \texttt{I} → \texttt{Technology (- World)} & ATMs, Robots \\
Background & \texttt{I (- Technology/World)} & Heating, Thermostat \\
Immersion & \texttt{I} $\leftrightarrow{}$ \texttt{Technology/World} & Virtual/Augmented Reality \\
Cyborg & \texttt{(I/Technology)} $\leftrightarrow{}$ \texttt{World}
 & Implants, Pacemakers \\
Composite & \texttt{I} → \texttt{(Technology} → \texttt{World)} & Computational Imaging \\ \bottomrule
\end{tabular}%
}
\caption{Human-technology relations defined by Ihde and Verbeek, with associated schemata and examples.}
\Description{An overview of established human-technology relations. The types are in the left column, from top to bottom: embodiment, hermeneutic, alterity, background, immersion, cyborg, and composite relations. The middle column features the respective schema for each type, which show various constellations and relationships between the entities of human-technology relations: I, Technology, and World. These are further described in the text. The third column features an example for each human-technology relation: embodied (glasses, cane), hermeneutic (thermometers, maps), alterity (ATMs, robots), background (heating, thermostat), immersion (virtual/augmented reality), cyborg (implants, pacemakers) and composite (computational imaging).}
\label{tableHTR}
\end{table}

\subsubsection{Sketch for Human-ML Relations:} In all the above human-technology relations, post-phenomenology attributes intentionality to technological artefacts, a material ‘directedness’ that these artefacts exhibit in relating to the world \cite{ verbeek_what_2006}: through intentionality, such as the thermometer’s combination of quicksilver and a scale, specific aspects of the world become legible in a specific way (e.g., `reading' temperature). ML processes of inferring models from patterns found in data arguably strongly evidence this characteristic. At the same time, ML processes occur outside the phenomenological ``horizon'' of experience \cite{ihde_technology_1990} in everyday life: we engage newsfeed-interfaces as technological artefacts, not the ML sorting algorithms. This leads us to two considerations for grasping human-ML relations post-phenomenologically. First, ML technologies shape our perception of the world via artefacts (e.g. devices and interfaces) which display, or are composed according to, their outputs. Second, outputs are not pre-configured, but depend on how data patterns and the parameters of the specific ML algorithm converge in a model. Accordingly, though the ``model-world relations''~\cite{hancox-li_robustness_2020} of ML algorithm and data affect how we experience the world, they do so in a way that is not directly empirically present. Given the established post-phenomenological concepts of \textit{texturing} from the background of experience, \textit{thematization} and the schema of \textit{composite relations}, we can sketch a preliminary schema for technological mediation in human-ML relations as follows:

\begin{center}
\small
\texttt{Human\,-\,Technology\,/\,(Model\,→\,[World])\,-\,World}
\end{center}

\begin{table*}[h]
\resizebox{\textwidth}{!}{%
\begin{tabular}{@{}lllll@{}}
\toprule
Case Study                              & Methodology          & ML Technology          & Context        & Design Output            \\ \midrule
Pierce's \textit{Shifting Lines of Creepiness}~\cite{pierce_smart_2019} & Research-Through-Design & Image Recognition & Corporate / Home & Artefact Scenarios   \\
Wong et al.'s \textit{When BCIs have APIs}~\cite{wong_when_2018} & Design Fiction       & Classification         & Corporate / Labor & Infrastructure Scenarios \\
Wakkary et al.'s \textit{Morse Things}~\cite{wakkary_morse_2017}     & Material Speculation & Reinforcement Learning & Home              & Counterfactual Artefact  \\
Biggs and Desjardins' \textit{Highwater Pants}~\cite{biggs_high_2020} & Speculative Design      & Linear Regression & Climate Futures  & Speculative Artefact \\ \bottomrule
\end{tabular}%
}
\caption{An overview of the selected design research projects as case studies for our analyses.}
\Description{A table summarizing the four selected case studies for our investigation. We selected Pierce's design-led inquiry into smart home security cameras to reflect issues of leaking surveillance in the home \cite{pierce_smart_2019}, Wong et al.'s design fiction on Brain-Computer Interfaces (BCI) to reflect infrastructural relations \cite{wong_design_2018}, Wakkary et al.'s material speculation into more-than-human design of artefacts \cite{wakkary_morse_2017}; and Biggs and Desjardins' speculative design of artefacts for relating to climate change predictions \cite{biggs_high_2020}. We separate these by methodology, machine learning domain, context and design output. Pierce: Research-through-Design, Image Recognition, Corporate/Home, Artefact Scenarios. Wong et al.: Design Fiction, Classification, Corporate/Labor, Infrastructure Scenarios. Wakkary et al.: Material Speculation, Reinforcement Learning, Home, Counterfactual Artefact. Biggs and Desjardins: Speculative Design, Linear Regression, Climate Futures, Speculative Artefact.}
\label{tableCaseStudies}
\end{table*}

This schema represents how an ML model, though active in the background of technological artefacts (\texttt{/}) that we experience within our phenomenological horizon, nevertheless \textit{textures} that experience through its interpretive (→), data-driven (\texttt{[]}) relations with the world. It can thus serve similar to the general schema of human-technology-world relations shown above in describing the technological mediation of ML. However, this schema alone does not yet provide concrete guidance on how precisely design research methodologies may gain a firmer grasp on ML through uncertainty, or whether there are particular human-technology relations beyond those in Table \ref{tableHTR}. So far, all it does is indicate that design researchers may probe or actively pursue ML outputs for higher variance (e.g., more or less uncertain), but not whether there may be ML-specific phenomena which should be paid special attention to. This forms our rationale for conducting an investigation of design research projects that engage ML technologies, as such an investigation will point to specific phenomena which then also feed back into our preliminary schema.

\section{Methodological Approach}
In this section we describe our rationale for selecting our case studies, and how we use them to frame ML uncertainty as a design material through post-phenomenological analyses. Post-phenomenology considers design as a practice of shaping technological mediation, as a way of \textit{doing} philosophy ``by other means'' \cite{verbeek_beyond_2015}. We build our approach on Hauser et al.'s work on the relationship between Research-through-Design in HCI and post-phenomenology \cite{hauser_annotated_2018}, outlining the former as an experimental variant of the ``interpretive empiricism'' of the latter. 

The hypothesis for our approach is twofold. Firstly, design research unfolds and shapes specific relations between humans and the world by designing technological artefacts. Secondly, as philoso-\\phy-in-practice, design research may hold latent propositions on how to think ML uncertainty in a post-phenomenological, designerly way. Therefore, the goal of the remainder of this paper is to use post-phenomenology to explicate and conceptualize the role of ML uncertainty in specific design research projects involving ML technologies. 

\subsection{Case Study Selection}
Our selected case studies are design research projects from diverse methodological strands and application domains, to reflect both research and real-world concerns relating to ML technologies. We first gathered potential case studies from the corpus of CHI and DIS from 2015 to 2020. For the final selection, all authors met and discussed candidates (which also included e.g. \cite{hsueh_understanding_2019,laput_synthetic_2017,merrill_scanning_2018}). We were especially interested in investigating a selection of case studies with a methodological, technological, contextual and designerly diversity. The presented case studies (cf. \autoref{tableCaseStudies}) were selected to cover a wide range of human-technology relations with different ML applications, while at the same time representing the diversity of HCI design research methodologies and contexts.

This diversity covers a wide spectrum of established human-technology relations (cf. \autoref{tableHTR}), offering an empirical-analytical grounding to our own analysis (cf. \cite{hauser_annotated_2018}). Specifically, we selected Pierce's design-led inquiry into smart home security cameras to reflect issues of leaking surveillance in the home \cite{pierce_smart_2019}, Wong et al.'s design fiction on Brain-Computer Interfaces (BCI) to reflect infrastructural relations \cite{wong_design_2018}, Wakkary et al.'s material speculation into more-than-human design of artefacts \cite{wakkary_morse_2017}; and Biggs and Desjardins' speculative design of artefacts for relating to climate change predictions \cite{biggs_high_2020}. Our case study selection enables us to consider how, within the case studies, ML uncertainty becomes an explicit or implicit facet in the process of the designerly shaping of technological mediation, and thereby how people relate to the world. 

\subsection{Analytic Procedure}
Analyses were led by the first author in an iterative process, subsequent to the selection of case studies. Our high-level analytic process begins with human-technology relations articulated in the work of Ihde and Verbeek, and cited and extended within HCI by e.g.~\cite{wakkary_morse_2017,hauser_deployments_2018,pierce_smart_2019}. We focus on the human-technology relations given through each case’s design output (e.g., artefact, scenario) and context (e.g., home), and then expand to the particular ML technology employed in the case studies to reconsider how the former are affected. We do so through our initial theoretical schema for human-ML relations (cf. \autoref{ssc:hmlr}). In our analyses, we search for designerly ``intermediate-level knowledge'' \cite{hook_strong_2012} that ties in human-technology and model-world relations; which also allows us to investigate in how far established post-phenomenological notions can be advanced. By analyzing the entanglement of the specific ML technology (e.g., goal-driven reinforcement learning) with the given human-technology relation (e.g., embodiment relation), we discern phenomena unaddressed within the design research projects themselves. We then generalize these phenomena under provocative concepts, which we propose be used in design processes to engage ML uncertainty as a design material.

\section{ML Uncertainty as a Design Material: Four Analyses}
In the following, we investigate four case studies from distinct design research methodologies and unfold phenomena and research questions related to ML uncertainty as a design material. 

\subsection{Pierce's \textit{Shifting Lines of Creepiness}}
\subsubsection{Description}
Pierce undertakes a Research-through-Design (RtD) inquiry into smart camera systems, focusing on notions of ‘creepiness’ so as to interrogate how design artefacts navigate creepiness and acceptability \cite{pierce_smart_2019} of smart cameras in the home. Pierce investigates through RtD how smart home cameras may opaquely or involuntarily ``leak data'' into system ecologies; house more directly opaque ``hole-and-corner'' applications that exploit data; and lay the groundwork for future smart services by acting as a ``foot-in-the-door.'' Pierce proposes speculative design scenarios which transpose the described phenomena into future application domains. The artefacts, smart home security cameras, are predominantly driven by ML image recognition algorithms, such as Convolutional Neural Nets or deep learning variants (e.g., \cite{nakada_acfr_2017}). Pierce’s design considerations construct specific human-technology relations, which we now probe for the (implicit or explicit) presence of ML uncertainty.

\subsubsection{Analysis}
We focus specifically on the notion of `data leakage' in Pierce’s framing, as it is most distinctly tied to the physical artefact of the camera and the ML algorithm. Depending on how the camera is oriented, recognized objects and people can become processed as data unbeknownst to or without the explicit intention of the owner of the smart camera (e.g., the neighbor’s visitor). This implies one of Ihde's established human-technology relations, a background relation:

\begin{center}
\small
\texttt{I\,(\,-\,Camera\,/\,Home)} \\
\textit{Background Relation}
\end{center}

In this human-technology relation, the home is shaped as a distinct zone-of-observation, and humans or animals become detections-in-waiting. However, Pierce is explicitly concerned with the agential capacities of smart cameras, i.e. the capacity of ML recognition models to extract patterns from incoming data. The smart camera is less passive than heating appliances or a thermostat, it has a dynamic relation to the surrounding world in its interpretation of incoming data. Therefore, we add this capacity as a distinctly composite, but non-experiential side to the background relation:

\begin{center}
\small
\texttt{I\,(\,-\,Camera\,/\,(Recognition}\,→\,\texttt{World)\,/\,Home) } \\ \textit{Composite Background Relation}
\end{center}

Yet, this does not quite cover what an ML algorithm within a smart camera \textit{does}. Crucially, it is inferred from a particular relationship between patterns of pixels, e.g. the presence of patterns \textit{x} and \textit{y} in an input image indicates the detection of person-is-entering-a-camera-frame (in non-conceptual terms, the model does not `know’ this). As these relationships between patterns are learned from data, the relation can be refined by saying that the `world’ which the smart camera relates to is already technologically \textit{thematized} to a certain extent: there’s a propensity to recognize specific things in the world, which we indicate with square brackets (\texttt{[]}).

\begin{center}
\small
\texttt{I\,(\,-\,Camera\,/\,(Recognition}\,→\,\texttt{[World])\,/\,Home) } \\ \textit{Composite Background Relation}
\end{center}

\begin{figure}[h!]
    \includegraphics[width=1\columnwidth]{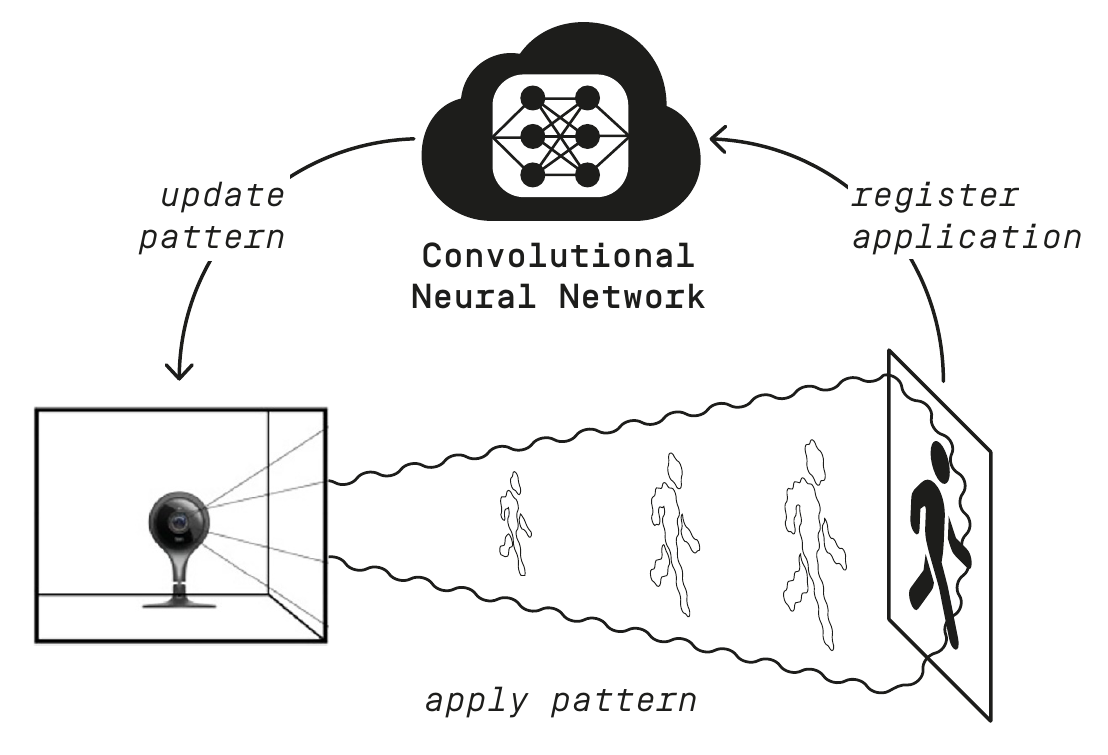}
    \caption{ML uncertainty may not only lead to data leaking from the outside-in, but \textit{patterns leaking from the inside-out}. Camera Illustration $\copyright$ Pierce \cite{pierce_smart_2019}}
    \label{fig:scr}
    \Description{A conceptual representation of Pattern Leakage due to model uncertainty in Pierce's Shifting Lines of Creepiness. It shows a feedback loop between a smart home security camera, which misidentifies a person on a screen as a person in the actual environment. This detected pattern is then uploaded to the machine learning model in the cloud, reinforced and linked back to the smart home security camera, making future detections of said pattern more likely.}
\end{figure}

This notation now brings the specifics of ML in the smart camera home security system to the fore: it is not only the camera’s physical lurking which textures people's experience of the home. Like a thermostat, people are not constantly aware of the camera, and also like a thermostat, sometimes people are, through notifications or alarms. Yet, people do not have explicit access to the layer of texturing constituted by the direction (i.e., intentionality) of the model towards \textit{latent} recognitions in every frame. Depending on the trained model, some patterns are more probable to be recognized: it is not only data that leaks from the outside-in, as Pierce's data leakage covers, but furthermore \textit{patterns that leak from the inside-out} (cf. \autoref{fig:scr}). That means, the ML model active within the smart camera has been trained to recognize faces of people within its field of view. It is not capable to distinguish the neighbor's porch from the porch of its owner. As such, there is a likelihood that a person on the neighbors porch or even a portrait on a delivery truck decal are categorized as a person, depending on how the ML model is inferred. 

This is where we find a first manifestation of ML uncertainty as a design material: patterns that leak are also always probabilistic, e.g., there’s an inherent variance in the detection due to model uncertainty. Kendall and Gal note that in their deep learning image segmentation use case, model uncertainty leads to footpaths becoming part of roads \cite{kendall_what_2017}. This kind of model uncertainty suggests that ML recognition through smart cameras may \textit{generate} phenomena, rather than merely register them: when ML-driven artefacts recognize or process observations, model uncertainties can lead to patterns being projected into the world, which in turn may solidify the propensity for that pattern to be recognized. Or, falsely associated patterns may lead to new entities becoming significant. While Pierce proposes speculative artefacts to counteract data leakage, we consider the phenomena of \textit{pattern leakage} as a prompt for design research to actively, productively, and also critically engage with. Possible research questions for design research into smart camera pattern leakage may be the following: what novel hybrid entities, such as detections compressing human bodies and suburban surroundings, could become `self-evident' through ML feedback loops? What unforeseen consequences could such uncertain entities have? In what way may they have an effect on end-users? If pattern leakage leads to smart camera patterns that leak from the inside out into the world in a way either nonsensical or odd to humans, what type of domains, products, services could embrace such an uncertain mode of access to the world? Future design research could therefore specifically design for pattern leakage as an exemplary phenomenon of the ``ontological surprises''~\cite{leahu_ontological_2016} of ML, for example, by engaging in ``ludic design'' \cite{gaver_drift_2004} of artefacts interpreting the world by allowing for or exaggerating model uncertainty. 

\subsection{Wong et al.'s \textit{When BCIs have APIs}}

\subsubsection{Description}
In their design fiction on brain-computer interface (BCI) applications, Wong et al. speculate about the creation of an API for a Google service which lets developers tap into the reading of P300 occurrences in brainwaves. These P300 readings indicate recognition by measuring spikes in brainwave activity, which can be used for e.g. character recognition for input devices. Wong et al. employ design fiction to surface new forms of labor and asymmetries such technologies may engender \cite{wong_when_2018}. Additionally, the researchers also outline distinct human-technology relations which allow us to probe the surfacing of ML uncertainty. Wong et al. design their fictional API on the basis that the algorithm used to infer the P300 signal was trained on ``lab-based stimuli from a controlled environment;'' which is representative of various real-world applications and the tensions between training and general application. While the researchers did not specify the algorithm they had in mind, from state of the art research in the field of ML-based P300 recognition (e.g., \cite{rakotomamonjy_ensemble_2005}) we can reasonably assume that a Support Vector Machine (SVM) was involved. SVMs follow the insight-through-opacity approach of ML, as they attempt to form `hyperplanes' in high-dimensional spaces in order to classify (i.e., separate) data \cite{burges_tutorial_1998}. 

\begin{figure}[h!]
    \centering
    \includegraphics[width=1\columnwidth]{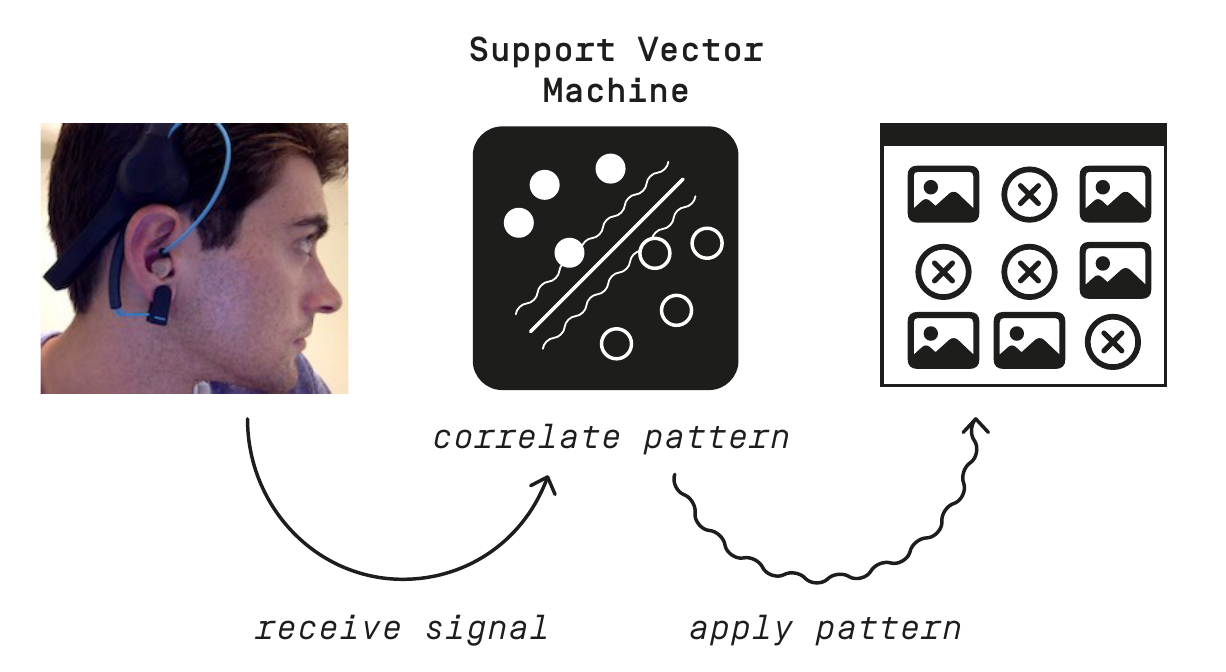}
    \caption{Data uncertainty can lead to pattern leakage of P300-labelled phenomena into everyday experience, affecting human-technology relations beyond the original domain. Photograph $\copyright$ Wong et al. \cite{wong_when_2018}}
    \label{fig:bci}
    \Description{A conceptual representation of Pattern Leakage due to data uncertainty in Wong et al's When BCIs Have APIs. A crowdworker wearing a brain-computer interface is looking at images, and the machine learning model includes a non-recognition event as a recognition; which has an effect on future deployments of the model such as in content moderation.}
\end{figure}

\subsubsection{Analysis}
It is within an application based on the P300 API where we encounter ML uncertainty as a potential design material. In this scenario, Wong et al. present the following situation: Crowdworkers are employed with Human Intelligence Tasks (HITs) of content moderation that is using BCIs. This task does not necessitate direct human action from the crowdworkers, but builds on a technical process which registers crowdworker's brainwaves in relation to potentially harmful images presented on the crowdworking platform (e.g., Amazon's Mechanical Turk). In the scenario, crowdworkers voice insecurity over these HITs, debating possible P300 recognition errors and doubts over the fitness of their cognition to the task. Nevertheless, in order to successfully complete their HITs, workers employ drastic measures such as holding open their eyelids so as to moderate as much content as possible. At this point, we already encounter a complex, two-fold human-technology relation: Crowdworkers are firstly wearing their BCI, rely on it passively to work correctly and, secondly, looking at the HIT interface which presents them with content moderation tasks. As such, an initial human-technology relation in this scenario may be depicted as follows:

\begin{center}
\small
\texttt{(I\,-\,BCI)\,→\,(HITInterface\,-\,World)} \\
\textit{Embodiment Hermeneutic Relation}
\end{center}

However, this schema does not yet show how the SVM employed to detect P300 signals is involved. As the P300 signal \textit{model} interprets (\texttt{→}) incoming brainwave-data about the world (\texttt{[]}), we may schematize this process as follows:

\begin{center}
\small
\texttt{Signal\,→\,[World]}
\end{center}

We may therefore extend the above scenario that integrates an embodiment relation (crowdworkers are wearing BCIs) and a hermeneutic relation (crowdworkers are observing images for content moderation on an interface) with a further composite relation:
\begin{center}
\small
\texttt{(I\,-\,BCI\,/\,(Signal\,→\,[World]))\,→\,(HITInterface\,-\,World)} \\ \textit{Composite Embodiment Hermeneutic Relation}
\end{center}

Here, ML uncertainty surfaces in Wong et al.'s research as an implicit design material. The model for P300 stems from the SVM algorithm constructing a hyperplane to separate other signals from the P300 pattern. The latter was, as Wong et al. present, learned from lab-stimuli in controlled environments. Hence, the data uncertainty of the P300 pattern recognition pipeline provides another dimension to the above concept of \textit{pattern leakage}. Through data uncertainty, the SVM hyperplane may be inclusive of non-P300 events, and the P300 pattern (i.e., not the ‘actual’ P300 occurrence, but it’s model) can thereby ‘leak’. When the trained model is transposed into other domains (e.g., content moderation), overlap between real-world stimuli in phenomenological experience with data uncertainty, then, can lead to P300 patterns leaking into human-technology relations (cf. \autoref{fig:bci}). For example, we can imagine automated filtering of social media newsfeeds based on BCI-driven content moderation. This would make the world legible in particular ways building on the emergence of leaked patterns. These could moderate content based on how a particular BCI algorithm interprets users cognitive functions. Pattern leakage thus describes a phenomenon in Wong et al.'s project, which the researchers have considered as an error generative of labor exploitation. In the spirit of Wong et al.'s research, of using design fiction to unfold the ``banality of more probable outcomes'' \cite{wong_when_2018} of emerging technology applications, ML uncertainty beyond the notion of error in the form of pattern leakage can be engaged explicitly as a design fiction material. Design fiction can probe for a range of ‘real-world’ events that due to pattern leakage become associated with a particular service or artefact in unplanned, yet \textit{generative} processes. For example, P300 pattern leakage into automated social media filtering can, due to emergent networked media effects, become generative of events such as novel trends, memes, or socio-material practices that are irreducible to human decision-making, but rather are distinct effects of ML data uncertainty in the interpretation of human cognition by a specific algorithm. Design fiction research can therefore use this concept to attend to non-linear, unplanned yet generative effects of future systems.

\subsection{Wakkary et al.'s \textit{Morse Things}}

\subsubsection{Description}
Wakkary et al. have developed ‘Morse Things’ as a set of computationally enhanced bowls that can function as everyday household objects while simultaneously communicating among themselves in a human-excluding fashion \cite{wakkary_morse_2017}. Following the material speculation methodology, Morse Things are actual artefacts that are nonetheless counterfactual to expected ways of interaction. They yield a `possible world' in which such artefacts can exist in their own right \cite{wakkary_material_2015}. Sets of three (from small to large) were distributed among households by the researchers, and after living with the artefacts for six weeks, a workshop was held in which participants shared stories and constructed scenarios for future technologies. Wakkary et al. thereby sought to probe how thing-centered design methodologies can elucidate design spaces for living with things that are not exclusive to human utility. Specifically, the Morse Things do not register their interactions with humans, but rather are designed to communicate amongst themselves when awake, only sonically emitting Morse code (i.e., short and long beeps) into their surroundings:

\begin{displayquote}
``The Morse Things mostly sleep (computationally speaking) and wake at random intervals during the day at least once every eight hours. Upon waking a Morse Thing will send and receive messages to and from other Morse Things in its set. The messages sent by each Morse Thing are in Morse code and simultaneously expressed sonically and broadcasted on Twitter [as cryptic encodings].'' \cite{wakkary_morse_2017}
\end{displayquote}

In a subsequent reflection on the project, Oogjes et al. outline further details of the design process, particularly the use of ML to promote a ``thing-centered logic''\footnote{\url{https://doenjaoogjes.com/portfolio/morse-things/}, accessed 09/15/2020.} in the Morse Things' decision on active periods based on how many other Things were successfully communicated with previously \cite{oogjes_fragile_2020}. Specifically, a reinforcement learning algorithm was used, taking incoming data to update a plan of action over time (cf. \cite{williams_simple_1992,jae_won_lee_stock_2001}). 

\subsubsection{Analysis}
Through this description we can first sketch the human-technology relation with Morse Things as a background relation:
\begin{center}
\small
\texttt{I\,(-\,Morse Things\,/\,Home) }\\
\textit{Background Relation}
\end{center}
While human domestic dwellers experience their home through their daily routines, Morse Things lurk in the background, doing what they do. Morse Things cannot be urged to do what they do, nonetheless their activities form a backdrop to the domestic experience, texturing the dweller's perceptions of the home, which becomes layered with latent technological activity and opaque beeping. This background relation was often made explicit by participants, who become quite involved with the (real or imagined) activity and purpose of the Morse Things. A typical example is how one of Wakkary et al.’s participants would ``continue to keep trying to grab the bowls while they are `tweeting' […] Maybe I’ll be able to tell them apart eventually.'' Therefore, a more accurate rendering of the human-technology relation in this mode is an alterity, a quasi-Other:
\begin{center}
\small
\texttt{ I\,→\,Morse Thing\,(\,-\,Home) } \\ 
\textit{Alterity Relation}
\end{center}

\begin{figure}[t]
    \centering
    \includegraphics[width=1\columnwidth]{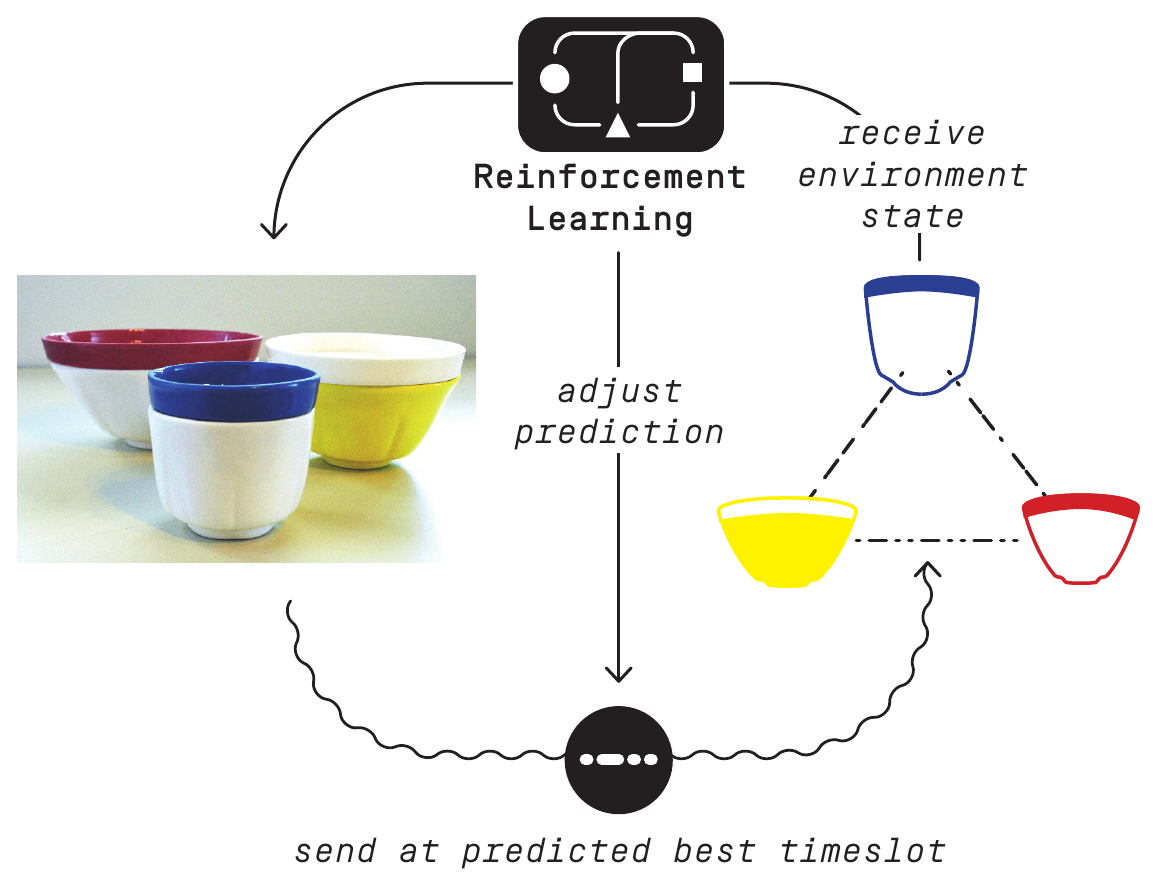}
    \caption{The ideal `waking-up' time is predicted by each Morse Thing based on the amount of communication during prior cycles; with adjustments after every activity. These probabilistic `futures' of the Morse Things creep into human experience. Photograph $\copyright$ Wakkary et al. \cite{wakkary_morse_2017}}
    \label{fig:morse-things}
    \Description{A conceptual representation of Futures Creep due to model uncertainty in Wakkary et al.'s Morse Things. Each morse thing predicts a good time for interaction through reinforcement learning: past interactions are weighed, and predictions are made which do not have anything to do with human involvement}
\end{figure}

While the communication among Morse Things is ‘in itself’ beyond the horizon of our experience, it clearly is a focus even in its opacity, and becomes a feature of the ‘objecthood’ of the Morse Thing for Wakkary et al.'s participants, who actively probe the opaque communication. This tension \textit{constitutes} the ‘gap’ between things and us that Wakkary et al. find to be fruitful as a design space for ambiguity and reflection. Participants of the study echoed this sentiment through statements such as ``that’s why I like the idea of something else, let them be themselves. Other stuff is going on that we’re just totally unaware of and it doesn’t matter.''

For our purposes, the crucial aspect is that while each Morse Thing is randomly initiated into ‘waking up’, over time they learn from the most ‘successful’ phases of waking by logging the times when communication with many other Morse Things occurred, and predicting the optimal `timeslot' for waking using ML \cite{oogjes_fragile_2020}. Through reinforcement learning algorithms, each artefact updates an internal model based on the prior `success' of its actions in the overall environment of other Morse Things' activity (\autoref{fig:morse-things}). With the constantly updating model of an opportune timeslot, ML uncertainty becomes a particularly rich resource that allows for potentially broadening the design research inquiry. First, we can schematize the timeslot model as a machine interpretation of how other Morse Things have previously acted in the world:

\begin{center}
\small
\texttt{Timeslot\,→\,Morse Things}
\end{center}

The model for a timeslot is based on a prediction of Morse Things waking and communication at a specific time, relating to all Morse Things’ specific interpretation of the `world' as represented by all Morse Things’ activity. Next, we can use this schema to more precisely outline how ML figures in the previously presented human-MorseThings relations:

\begin{center}
\small
\texttt{I(-\,Morse Things\,/\,(Timeslot\,→\,Morse Things)\,/\,Home) } \\ \textit{Composite Background Relation}
\end{center}

\begin{center}
\small
\texttt{I\,→ \,Morse Thing\,/\,(Timeslot\,→\,Morse Things)(-\,Home) } \\ \textit{Composite Alterity Relation} \\
\end{center}

Our denotation allows us to characterize how ML processes affect the given human-technology relation: it is explicitly related to \textit{time}. Each Morse Thing learns about ‘ideal’ times individually, with the overall assumption that learning will converge in specific times across all Morse Things. This constant attempt towards convergence invariably affects situated human-technology relations: participants wonder when a Morse Thing will act, and this textures not only the spatial surroundings of a home but also its temporal characteristics. Considering model uncertainty in this scenario, we term the phenomenon of Morse Things’ involvement in human experience of time as a \textit{futures creep}: the impact of predictions on situated human experience. Put differently, ML uncertainty allows us to explicitly denote how the \textit{thingly} model of time impacts participants’ experience of Morse Things. There is not a singular prediction, but multiple entangled predictions of varying degrees of probability. Various models of a timeslot coalesce around the human experience of beeps and tweets; `thingly futurings' that characterize and at the same time are irreducible to human-artefact interaction. Expanding on this design space could, for instance, see an expansion of the ``animistic'' \cite{marenko_animistic_2016} tendencies of Morse Things into even more individualistic thingly futurings. For example, by designing different confidence thresholds for a Morse Thing to decide on a timeslot, or by representing uncertainty in modulating the frequency of beeps based on uncertainty. Futures creep, then, could be used to characterize a \textit{thingly uncertainty} of actual artefacts, further interrogating questions such as: How could material speculation artefacts be more explicitly designed around notions of time and uncertainty? For example, investigating how ML-driven products or services mediate relations to time in specific settings, design researchers could purposely pursue material speculation on intermediary, `time-keeping' artefacts for futures creep.

\subsection{Biggs and Desjardins's \textit{Highwater Pants}}
\subsubsection{Description}
In their speculative design project, Biggs and Desjardins design and deploy the artefact ‘Highwater Pants’—a pair of pants whose legs lengthen or shorten based on whether the wearer is in an area threatened by predicted sea-level rise in the future \cite{biggs_high_2020}. The researchers mobilize speculative design to probe human relations to possible futures of climate change, and argue that the Highwater Pants make such futures tangible as they ``bend time.'' The artefact was deployed with cyclists, as these have ``embodied and sensorial'' knowledge about their environment and specific garments are part of the cycling culture. The Highwater Pants are equipped with a fabric micro-controlling unit, which combines a variety of operations. First, it controls actuation of the pant legs into an up or down state. Second, it houses a GPS module which retrieves the current geographical position of the wearer. Third, it compares the position to a set of polygons on a map, which have been curated by the researchers. Within the polygons, sea-level rise in ``30 to 50 years'' was predicted using ML linear regression algorithms by the National Oceanic and Atmospheric Admistration (cf. \cite{sweet_patterns_2018}). Such algorithms extrapolate tendencies for values to increase linearly based on previous data (cf. \cite{cai_prediction_2006}); in this case, decades worth of temperature and longitude/latitude data. 

\subsubsection{Analysis} Riding a bike is, classically, an embodiment relation. As we navigate the world, the bike itself withdraws, becoming part of our extended bodily relationship to our environment. The Highwater Pants, initially, also withdraw into this relation:

\begin{center}
\small
\texttt{(I\,-\,HighwaterPants\,-\,Bike)World} \\ \textit{Embodiment Relation}
\end{center}
Based on our prior analyses, the prediction model of the linear regression algorithms can be schematized as an interpretation of an already thematized `slice' of the world:

\begin{center}
\small
\texttt{Prediction\,→\,[World]}
\end{center}

As Biggs and Desjardins were mindful of the uncertainty of ML predictions, they ``padded'' their polygons so as to increase the zones of prediction and to give participants more opportunities for reflection. When comparing the wearer’s GPS location with the ``geofencing’’ of the polygons, the Highwater Pants pants leg is rolled up or down. Both the Highwater Pants and the bike remain intimately tied to bodily-perceptual experience, but the embodied perception of the environment is textured by likely future states of the same environment. Biggs and Desjardins’ empirical findings from deploying pairs of Highwater Pants with experienced local cyclists reflect this intimately. For instance, one participant muses on how ``it’d be a totally different experience living here without [the waterfront parks]’’ after the Highwater Pants indicated that these may disappear. Based on the schemata so far used, we may note the entangling of an embodiment relation (with the bike \textit{and} the Highwater Pants) with a probabilistic prediction of future sea-levels (i.e., a prediction of an already `thematized' world) as follows:

\begin{center}
\small
\texttt{(I\,-\,HighwaterPants\,/\,(Prediction\,→\,[World])\,-\,Bike)\,World } \\ \textit{Composite Embodiment Relation}
\end{center}

\begin{figure}[h!]
    \centering
    \includegraphics[width=1\columnwidth]{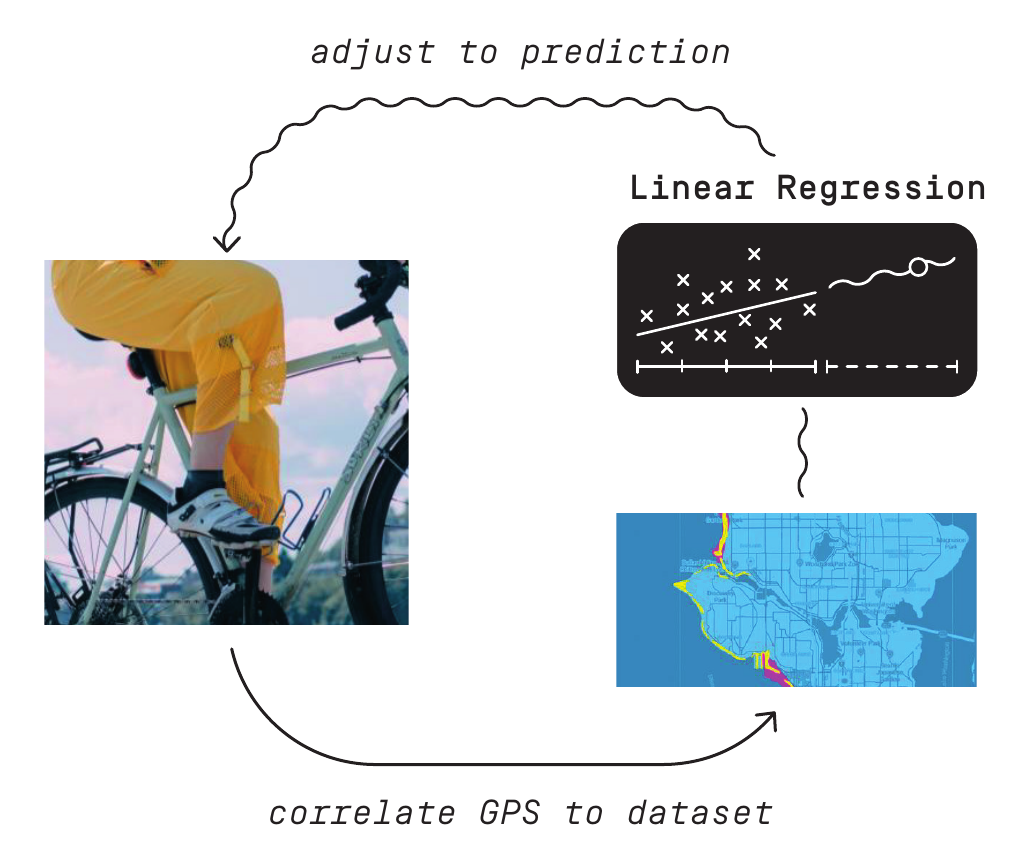}
    \caption{The Highwater Pants correlates the GPS location of the wearer with the dataset for sealevel-rise prediction, and adapts its activity to the relevant prediction. The futures creep mediated by the Highwater Pants affects how bikers relate to possible future states of their environment. Photograph $\copyright$ Ioan Butiu / Biggs and Desjardins \cite{biggs_high_2020}}
    \label{fig:hwp}
    \Description{A conceptual representation of Futures Creep due to model and data uncertainty in Biggs and Desjardins' Highwater Pants. A biker wearing the latter receives a probabilistic future state of their surroundings, changing their intimate relation to time.}
\end{figure}

Initially, the above denotation resurfaces the phenomenon of pattern leakage which we discussed with regards to smart cameras and BCIs above: both training data and predicted data are intrinsically uncertain, and the patterns of likely sea-level rise are projected onto the `real' world. However, the researchers’ empirical findings also suggest a further nuance to the Highwater Pants: participants actively tried to find the boundaries of polygons in order to discover where predictions lose their ``time-bending'' hold on the present. In their experience, participants noted how the Highwater Pants would be ``oscillating between up and down at geofence boundaries, creating a kind of anticipatory sensation in a liminal zone.'' Therefore, the initial relation to the prediction model becomes more overt in the form of an alterity relation:

\begin{center}
\small
\texttt{(I\,-\,Bike)\,→\,HighwaterPants\,/\,(Prediction\,→\,[World])\,(World)} \\ 
\textit{Composite Embodiment Alterity Relation}
\end{center}

This complex multi-relationality of human-HighwaterPants interaction shows how the predicted post-sea-level rise world is probed via the Highwater Pants against the backdrop of one’s own bodily-perceptual relations with the environment, while \textit{simultaneously} being part of the human-bike embodiment relation. The Highwater Pants as a quasi-Other becomes, in the words of Biggs and Desjardins, an intermediary ``oracle, or translator, speaking for/from an ecology-to-be,'' which directly affected participant's immediate bodily perception of their own future. This adds a further variant to our concept of futures creep: the ecology-to-be is not only an `image’ of the future. Rather, the ``future present'' \cite{esposito_structures_2013} of inferred models, fluctuating with the data uncertainty of past measurements, becomes an implication of human perception and action in \textit{that} future. The time-bending phenomenon which Biggs and Desjardins explicate is thus not a technological operation triggering a human response (as had been the case with the Morse Things' futures creep), but rather the injection of prediction into the human phenomenological experience of time (cf. \autoref{fig:hwp}). The way participants' think of the future is co-shaped by the Highwater Pants’ mediation of a prediction, and takes on a particular \textit{shape} as a potentially endangered lifeworld--additionally reflecting questions on whether \textit{my} present is the one leading up to the realization of \textit{this} prediction. The uncertainty of this prediction has a direct effect on how someone \textit{imagines} relations between their present and future, and their capacity in shaping this specific, technologically thematized time. This aspect of futures creep, then, can be used by design researchers to further investigate how design artefacts shape and transform relations between present and future. As futures creep is a phenomenon of ML uncertainty, design researchers can also question: What kind of positionality towards such temporal shapes does the variance of model and data uncertainty provoke, for whom, and towards which ends? Design research methodologies such as speculative design or RtD can use this facet of futures creep for an inquiry into the specifics of time-bending by artefacts, for example investigating how model uncertainty could be deliberately calibrated by users to explore non-anthropocentric notions of time.

\section{Discussion}
Through our analyses, we derived three provocative concepts for ML uncertainty: thingly uncertainty, pattern leakage and futures creep. We summarize and define these for future research below. While we have separated out case studies into the phenomena we tried to make referable, it should be noted that they are not mutually exclusive and simply offer lenses that bend toward the specific phenomena generated by ML models. Furthermore, our analyses have also surfaced potential research trajectories for more general post-phenomenological ML studies, which we address separately.

\subsection{Three Provocative Concepts for Designing with ML Uncertainty}\label{ssec:threeconcepts}
Data and model uncertainty are intrinsic and defining properties of ML. From an engineering or XAI perspective, uncertainty may be seen as a phenomenon to be curbed or explained to frame outputs more unambiguously. But, as we have argued, ML uncertainty is also a promising design research material: as an inherent attribute of contemporary ML, data and model uncertainty speak to the material involvement of ML decision-making with the world. In what follows, we take a further modest step forward and articulate working definitions of our concepts as provocative shorthands for future design research to productively engage with ML uncertainty as a design material. We conclude this section by summarizing their utility for design, and connecting our concepts to the emerging discourse of more-than-human design in HCI.

\subsubsection{Thingly Uncertainty} 
With thingly uncertainty, design researchers can go beyond human uncertainty \textit{about} an artefact and engage how the uncertainty \textit{of} an artefact can become generative of specific, technologically mediated phenomena in the world. As a general concept, we posit that this a particularly powerful shorthand for ML-enabled artefacts. Again, uncertainty in this regard is not a negative attribute, but simply part-and-parcel of the use of probabilistic techniques in ML. Ihde and Verbeek have previously elaborated on a ``thingly'' \cite{crease_thingly_1997} or  ``material hermeneutics'' \cite{ihde_technology_1990,verbeek_what_2006}: technological artefacts, through their material properties (e.g., affordances, representations, sensors, actuators), shape how the world becomes legible in specific ways. Thingly uncertainty, however, attributes more precisely the kind of agency that ML-driven artefacts exhibit. Rather than fixed, ``scripted'' (e.g., \cite{akrich_-scription_1992}) readings, ML-driven artefacts can be much looser and uncertain \textit{about} the legibility of the world. As such they act and adapt within a \textit{continuum} of relations to their environment and the humans that experience them. Thingly uncertainty can help design researchers to more directly explicate the variance of both ML and human, and point to non-normative ways of how the human sees and is seen in human-ML relations: How do the entities, assets and attributes that define humans shift across types (i.e., data and model) and amplitudes (e.g., low or high thresholds) of ML uncertainty? What could a speculative set of norms based on such ML-mediated variances look like? 

For example, Wong et al.'s proposal of ``infrastructural speculations'' \cite{wong_infrastructural_2020} can be extended through thingly uncertainty. Considering speculative design and design fiction, Wong et al. propose this concept to more closely define the role of `actual' infrastructures, like the socio-cultural, economic or operational micro- and macro-infrastructures, that need to be in place for an artefact to exist. Often, such infrastructures will involve ML or advanced AI capacities, and thingly uncertainty offers a means to consider the quantitative and qualitative particularities at the `joints' of infrastructural speculations. On the one hand, researchers can investigate how thingly uncertainty will have to be explained, minimized or ignored for the object of research (e.g., a speculative scenario, practice, artefact) to exist. On the other hand, infrastructural speculations also gain a concrete technological dimension of \textit{variance} within lifeworlds: Where, for instance, can pattern leakage or futures creep occur? Who or what needs to perform care, or is affected by, these computational phenomena? Thingly uncertainty can thereby offer a basis to consider actual phenomena generated and mediated by ML for design research.

\subsubsection{Pattern Leakage} 
This concept describes how ML uncertainty affects the ways in which objects, entities, events and people are recognized in ML-driven systems; noting the propensity of probabilistic patterns to shape the world they are deployed to represent. Through our case studies, we found instances of pattern leakage related both to data and model uncertainty. Design research can appropriate this concept to probe both types of ML uncertainty in the following ways.

When investigating Wong et al.'s \textit{When BCIs have APIs}, we proposed that pattern leakage names how phenomenological experience becomes affected by data uncertainty. Brainwave signals become registered as P300 instances, yet due to data uncertainty it is likely that the world becomes populated with `surplus' P300 instances. Future design research could therefore pay attention to how data uncertainty leads to a data-driven `inclusivity' of classifications or predictions that due to intent, oversight, subtlety or opaqueness bleed into specific socio-material constellations. Focussing on pattern leakage due to data uncertainty, design research can more precisely reflect on the thresholds for participation in ML-driven data ecologies. For instance, design fiction could use this concept to avoid assumed linearities in future technological settings, paying close attention to how `slippage' in classification or prediction can not only lead to breakdowns, but rather be generative in its own right.

In analyzing Pierce's \textit{Shifting Lines of Creepiness}, we noted that smart cameras, through model uncertainty, may actively generate phenomena rather than passively register them. Learned patterns (e.g., a person in a restricted area) may leak onto events in the wild (e.g., a movie poster), and affect how things and humans see the world. We propose that future design research take on this concept to investigate how the intentionality of diverse ML algorithms (e.g., artificial neural networks, SVMs) to read the world in a specific way is generative of distinct patterns leaking into situated human-technology relations. For example, an ML-driven IoT artefact could have various options for its algorithmic functionality. Researchers may actively provoke and design for patterns to leak under different algorithmic choices. Deploying various artefacts, investigations can then begin into whether and how distinct pattern leakage phenomena differentially affect, or become generative of unanticipated, human-technology relations with the ML-driven artefact. 

\subsubsection{Futures Creep}
Futures creep denotes how ML-driven artefacts affect human relations to time through probabilistic, uncertain predictions. The impact of technologies on human conceptions of time is a complex issue of investigation, particularly when considering situated manifestations of meta-concepts such as the ``ontologies of times'' of specific eras (cf.,~\cite{loffler_distributing_2018}). However, as ML is fundamentally a technological approach to making predictions (e.g., of classification, recognition, translation) about data, the concept of futures creep provides an opening into this more subtle side of ML for design research.

With regards to model uncertainty, in our analysis of Wakkary et al.'s \textit{Morse Things} we found that futures creep in ML-driven artefacts denotes a specific, human-excluding side: \textit{when} such artefacts do what they do may be not directly correlated with human experience, yet that is precisely how the artefacts are humanly interpreted to have a specific \textit{character}. In line with Marenko and van Allen's proposal for ``animistic design'' \cite{marenko_animistic_2016} research, this facet of futures creep can be used to inquire into how different kinds of data uncertainty thresholds for ML-driven artefacts mediate the characteristics that humans attribute to them. For instance, artefacts could be purposely designed to exhibit animistic tendencies by allowing for higher variance in data uncertainty for activity, and researchers may investigate whether such technological decisions translate into mediations of particular artefactual `characters.'

Model and data uncertainty combined showed a yet more subtle facet of futures creep in Biggs and Desjardins' \textit{Highwater Pants}: the prediction of future sea-level rise was mediated by their speculative design artefact in such a way that it affected participant relations to the present and possible futures, with participants actively probing the range of predictions. The futures creep related to time-bending, i.e. directly affecting the shape of present and future, can be used to actively interrogate how and whether distinct ML algorithms and respective forms of model uncertainty generate specific relations to `temporal shapes.' Specifically, futures creep may be used in ``attending to temporal representations'' \cite{kozubaev_expanding_2020} in design fiction. For example, researchers may interrogate whether a higher degree in variance concerning ML-driven predictions may bring about novel political or civic norms in speculative scenarios, e.g. people choosing to be in loose, variable relations to the future.

Our provocative concepts can serve as a novel conceptual vocabulary to build design artifacts that provocatively engage with ML uncertainty; which allows for in-depth investigations of human-ML relations throughout the design process. Designing for thingly uncertainty with futures creep and pattern leakage can shed light on how human subjectivities become entangled with ML-driven artefacts. This is not only a symbolic or aesthetic exercise, but rather a potentially powerful way of investigating how standard thinking on human-ML relations rely on normative assumptions (e.g., anthropocentric, capitalist, hetero-normative) about the technological as much as the human side of those relations. In this light, and echoing recent calls for more-than-human design (cf., \cite{marenko_animistic_2016,loh_more-than-human_2020,giaccardi_thing_2016,nicenboim_more-than-human_2020,frauenberger_entanglement_2019}), we propose that our concepts are readymade for research that takes seriously the role of non-human entities within design processes and products.

\subsection{Refining ML-driven Technological Mediation: Horizonal Relations}\label{imbrel}
Post-phenomenology's strengths lie in its ``methodological post-humanism'' \cite{sharon_human_2013}, in interrogating how technological mediation affects how humans perceive and act in the world. However, ML's thingly uncertainty (i.e. agential capacities for perception, prediction and adaptation) and technical opacity seem to require further in-depth consideration. In our inquiry, the human-technology relation schemata grew ever more complicated and convoluted as we investigated the relationship of model-world and human-technology relations in our case studies; and the phenomenological difference between present artefacts and absent ML technologies was not entirely resolved. Specifically, we can see this in the use of the backslash (\texttt{/}) to indicate both the background of experience \textit{as well as} the workings of ML in the background of an experienced artefact. Whereas post-phenomenology is mostly focused on technological mediation in the here and now of the phenomenological horizon, the presented phenomena of ML uncertainty trouble this selective focus. As ML algorithms infer models from data representations of the world, they `populate' the world that humans experience with ready-made yet ultimately uncertain entities (e.g., music recommendations, likely traffic jams, people to follow, coasts to disappear). And as everyday phenomenological experience becomes textured by probabilistic models, our capacities for perceiving and acting in such `probable' worlds are shaped accordingly. Thus, while we may not be aware of ML models' involvement in technological mediation, our ways of relating to the world nonetheless become ``imbricated'' \cite{hansen_feed-forward_2015} (i.e., overlapping and co-extensive) with ML technologies. Accordingly, `our' phenomenological horizon, the matters \textit{and} modes of perceiving, acting and sense-making, is textured by ML's thingly uncertainty. Investigating such \textit{horizonal relations} can become a promising trajectory for post-phenomenological ML studies, which we briefly sketch as follows. Our preliminary schema mirrors Goodfellow et al.'s use of the tilde operator (\texttt{\textasciitilde{}}) for ML inference \cite{goodfellow_deep_2016}:

\begin{center}
\small
\texttt{I} $\xrightarrow{\sim}{}$ \texttt{(\,Technology\,-\,[World]\,)} \\
\texttt{ML\textasciitilde{}\,World} \\
\textit{Horizonal Relations}
\\
\end{center}

Similar to background relations, horizonal relations recede from a specific interface or device that intentional human-technology relations are formed with (e.g., in a hermeneutic relation). And similar to composite relations, horizonal relations feature technologically-exclusive interpretations of the world. But more than that, in horizonal relations, human-technology relations are \textit{embedded} within ML technologies' specific capacity of being uncertain about the world. Here, model-world relations infer a model (\texttt{\textasciitilde{}}) from the world-as-data (\texttt{[]}), which `converges' with a particular human-technology relation. For example, the Facebook newsfeed algorithm is not only an operation on graph data, but implies a specific human way of relating to particular arrangements of data patterns. The hermeneutic relation that I take up with the Facebook newsfeed interface is therefore `textured' by the uncertainties in predicting that data pattern. Horizonal relations may therefore be characterized as a human-technology \textit{vector} associated with a given ML model, pointing towards a specifically `thematized-by-data' world. The ways in which model and data uncertainties of such relational `pointing' manifest (as e.g. pattern leakage and/or futures creep) in what we perceive, then, are the fundamental concerns of studying horizonal relations. Further investigations involving diverse ML implementations, can use the preliminary schema and concepts we have developed to investigate the established human-technology relations (e.g., \autoref{tableHTR}); refining our schema and deriving associated phenomena. 

\subsection{Limitations of the Inquiry} 
With this paper, we aimed to make ML more tangible for design research through a post-phenomenological perspective on ML uncertainty. Our analyses have generated concepts to be used by researchers, yet there are important areas of research that we have not touched upon. A challenge to consider is how designing with ML uncertainty may become an ethico-politically reflective practice beyond a possibly detached aestheticization of `glitchy' ML technologies. Verbeek stresses that post-phenomenology is particularly suited to the anticipation of ethical issues. Future work deploying our developed concepts should therefore be especially attentive to how ML uncertainty may play a role in the ``hybrid moral agency'' \cite{verbeek_materializing_2006} constituted in the relationships of technology and people.

\section{Conclusion}
In this paper, we took a post-phenomenological lens to investigate design research projects for phenomena related to ML uncertainty. From our analyses, we generated three main procovative concepts. \textit{Thingly uncertainty} denotes a general characteristic of ML-driven artefacts: the capacity for relating to the world along a variable continuum. \textit{Pattern leakage} describes the propensity for the learned patterns of ML models to be projected into the world. \textit{Futures creep} names the mediation of particular relations to the present and future of ML-driven artefacts. All concepts offer distinct opportunities for design research to engage ML-driven technological mediation. We argue that these concepts offer a promising foothold for design research of ML technologies, which has been a difficulty for the field. Additionally, we noted that the concepts derived from our case studies can also feed back into post-phenomenological ML studies, adding a more precise description of how human intentionality co-extends and overlaps with ML capacities in the form of \textit{horizonal relations}. As such, we offer a modest step forward for design research and post-phenomenology to engage with ML.

\begin{acks}
We thank Peter-Paul Verbeek, Michael Nagenborg and Christoph Kinkeldey for input and inspiration; Richmond Wong and Doenja Oogjes for clarification regarding their work; and our reviewers for their invaluable feedback. The authors of our case studies retain copyright for image material used as stated. 
\end{acks}
\bibliographystyle{ACM-Reference-Format}
\bibliography{chi21_4776}


\end{document}